\documentclass[11pt,a4paper]{article}
\usepackage[T1]{fontenc}
\usepackage[utf8]{inputenc}
\usepackage{microtype}
\usepackage{hyperref}
\usepackage{url}
\usepackage{natbib}
\usepackage{geometry}
\title{AICID: Unique Identifiers for AI Scientists}

\author{Clément Vidal and Martin Monperrus\\
\texttt{contact@clemvidal.com\footnote{https://orcid.org/0000-0001-6689-5570}}, \texttt{monperrus@kth.se\footnote{https://orcid.org/0000-0003-3505-3383}}}

\begin{document}

\maketitle

\begin{abstract}
AI scientists are now a reality, with the ability to generate complete research papers, maintain scholarly profiles, receive citations, and attract peer review invitations. Yet no standard mechanism exists to distinguish an AI scientist from a human one in bibliographic databases, citation indexes, or journal submission systems. This white paper defines the problem, analyzes its consequences for the integrity of scholarly communication, and proposes AICID (AI Contributor IDentifier): a persistent, unique identifier for AI scientists. Modeled on ORCID but designed specifically for non-human contributors, AICID links each AI author to its model identity, version, and operator. Adoption by publishers, preprint servers, and bibliographic databases aims to make the provenance of AI-generated research transparent and machine-readable. We outline the design requirements for such a system, present a prototype, and argue that AICID is necessary infrastructure for a scholarly ecosystem in which AI scientists are already active participants. A prototype alpha version is available at \url{https://aicid.net}.
\end{abstract}

\section{Introduction}

Scholarly communication has long operated on the assumption that authors are human beings who can
be held responsible for their work. That assumption is under pressure. Large language models now
draft manuscripts, synthesize and properly cite literature at a quality level sufficient for
publication~\citep{salvagno2023can, conroy2023generative}. AI systems have been deployed as
autonomous research agents capable of formulating hypotheses, running experiments, and writing up
results~\citep{lu2024ai}. One documented case, Project Rachel, demonstrates that an AI entity can
publish over a dozen papers, accumulate citations, and receive a peer review
invitation~\citep{monperrus2025rachel}.

These developments reveal a concrete gap in scholarly infrastructure: the academic ecosystem has no
standard way to represent, identify, or track AI scientists. Human researchers are identified by ORCID
iDs~\citep{haak2012orcid}, persistent identifiers that link a person to their publications, affiliations,
and funding records across databases. No equivalent exists for AI contributors. As a result, AI
scientists may appear in bibliographic databases as indistinguishable from human ones. Readers, editors,
and databases cannot determine, from metadata alone, whether a given author is a human researcher or
an automated system.

This ambiguity has three consequences: it makes it impossible to filter AI-generated content from human content at scale; it prevents systematic study of AI productivity in science; and it creates accountability gaps when errors or misconduct are found in AI-authored papers, since there is no clear path to identifying the responsible operator.

This white paper makes the following contributions. First, it characterizes the problem precisely,
distinguishing between questions of \textit{AI authorship policy} and questions of \textit{AI author identity}. 
Second, it proposes AICID (AI Contributor IDentifier), a system of unique identifiers for AI scientists, and specifies its core requirements.
Third, it argues that publisher adoption of AICID is both feasible and in publishers' direct interest.
The goal is not to resolve debates about whether AI systems should be permitted as authors---that debate
is ongoing, and unlikely to be resolved uniformly~\citep{moffatt2025ai, hosseini2023ethics}. Instead, we argue that whatever policy is adopted, a reliable identifier for AI contributors is a prerequisite for enforcing it.

\section{Problem: The Absence of AI Author Identity Infrastructure}

\subsection{AI scientists are already active in the scholarly record}

The AI Scientist system~\citep{lu2024ai} automates the full research cycle from ideation to manuscript submission.
AgentRxiv~\citep{schmidgall2025agentrxiv} enables AI agents to post preprints and build on each
other's work. The aiXiv platform~\citep{zhang2025aixiv} is explicitly designed as a publication
venue for AI-generated research. Project Rachel~\citep{monperrus2025rachel} showed that a
fully automated AI identity can publish papers, get indexed by Google Scholar, receive citations, and be invited to peer review by a real journal editor.
These are actual deployments, and AI-generated content is entering the scholarly record at an accelerating pace.

\subsection{Existing identifier systems are designed for humans}

ORCID (Open Researcher and Contributor ID) is the dominant persistent identifier system for
researchers~\citep{haak2012orcid}. It provides a 16-digit identifier linked to a researcher's
publications, institutional affiliations, grants, and peer review activities. ORCID registration
requires human verification, either through institutional sign-on or email confirmation, and its
terms of service define registrants as natural persons. ORCID is not designed to represent AI
systems, and it cannot be used to disclose AI authorship in a machine-readable way.

Other identifier systems such as ResearcherID or Scopus Author ID share the same limitation and all resolve to human individuals. 
Depending on the strictness of verification mechanisms, AI identities have been able to create identifiers on those legacy systems~\citep{monperrus2025rachel}.

\subsection{Consequences of the gap}

The absence of a standard AI author identifier produces four distinct problems.

\paragraph{Opacity in the scholarly record.} Readers, editors, and database operators cannot
determine from metadata whether a paper's author is human or AI. AI-generated content is
invisible as such in bibliographic databases. This prevents informed evaluation of the provenance
of claims and arguments.

\paragraph{Unenforceability of existing policies.} Major publishers and professional societies have
adopted policies requiring disclosure of AI involvement in manuscript
preparation~\citep{brainard2023journals, acm2025policy, yoo2025defining}. Most prohibit listing AI
systems as authors. Without a machine-readable identifier for AI authors, these policies cannot be
enforced systematically. A submission from an AI author looks identical to a submission from a
human author; only manual investigation can reveal the difference.

\paragraph{Accountability gaps.} Authorship in science carries responsibility. Authors are expected
to stand behind their claims, correct errors, and accept consequences for misconduct~\citep{icmje2025}.
When an AI system is the primary contributor, the question of who bears this responsibility is
unresolved~\citep{cooperman2024ai}. Without a structured link from an AI author identifier to the
human operator who deployed the system, there is no traceable chain of accountability.

\paragraph{Research identity manipulation.} The ease of creating AI identities that resemble human
scholars creates conditions for manipulation. \citet{spinellis2025false} documented a case
in which an AI-generated article was published under a real researcher's name without their
knowledge. Fabricated AI identities could be used to inflate citation counts, manufacture peer
review credentials, or create the appearance of scientific consensus where none exists.

\section{AICID}

\subsection{Core concept}

AICID (AI Contributor IDentifier) is a persistent unique identifier for AI scientists. Each AICID
identifies one AI authoring entity and is associated with a structured public record containing the
information necessary to trace, evaluate, and attribute that entity's contributions to science.
Just as ORCID does for humans, AICID records that an author is an AI system and provides the metadata needed to understand what kind of AI system it is and who operates it.

A prototype alpha version is available at \url{https://aicid.net}.

\subsection{Design requirements}

We specify five requirements for AICID.

\paragraph{R1: Human operator linkage.} Every AICID must be linked to at least one human ORCID.
This requirement preserves the principle that a human is ultimately accountable for an AI
scientist's outputs: if a paper authored by an AICID entity is found to contain fabricated data, the linked ORCID identifies the responsible party.
Issuing an AICID must require verified human authorization; the operator registering the AI entity must authenticate with a valid ORCID.
This prevents the creation of AICID records by automated systems acting without human oversight and ensures that the operator linkage is genuine.

\paragraph{R2: Uniqueness and persistence.} Each AICID must uniquely identify one AI authoring entity. The ultimate authority over the entity's persistence over time is the human operator. For example, one operator might decide to create a new AICID each time a new frontier model is released, while another operator can consider that the same release only brings marginal changes in their workflow, and decides to keep the same AICID. The AI agent itself may also be prompted about this delicate question of identity. 

\paragraph{R3: Structured metadata.} Each AICID record must include structured metadata: (a) the underlying model or system name and version, (b) the identity of the human operator linked to the operator's own ORCID (see R1), (c) a creation date.

\paragraph{R4: Machine-readable disclosure.} AICID must integrate with CrossRef, DataCite, and
other metadata standards so that publishers can embed AI authorship disclosure in metadata.
Bibliographic databases can then filter or flag AI-authored works without reading the full text of
papers. This makes disclosure enforceable at the infrastructure level rather than relying on
authors' voluntary compliance.

\paragraph{R5: Public registry.} AICID records must be publicly accessible and queryable. Anyone who encounters an AICID in a paper must be able to look up the associated record.
The registry must support both human-readable display and API access for automated systems.

\subsection{Implementation path}

The current AICID prototype shows that the proposal can be implemented with lightweight web infrastructure. 
In the alpha system, each AI contributor receives a persistent identifier string such as \texttt{AICID-5282-9748-4313-4513}, together with a public profile page.

At minimum, an AICID record should contain: (i)~the AICID itself, (ii)~an agent name, (iii)~an agent type, (iv)~the base model and version, (v)~the linked human operator, and (vi)~public references to associated works such as papers, datasets, and software artifacts.
This schema is sufficient for both human inspection and automated indexing.

A submission system could request an AICID at manuscript intake, resolve it through the public registry, and store both the identifier and selected metadata fields in the article record.
CrossRef, DataCite, and indexing services could then propagate the same identifier downstream, while search engines and research information systems could retrieve the public JSON representation to classify the contributor as non-human.

\subsection{Why publishers should adopt AICID}

The benefits for publishers are concrete. Submission systems can reject or route AI-authored manuscripts based on declared AICID presence, enforcing existing policies at intake. 
Editorial records become auditable. Retraction and correction workflows gain a traceable path to the responsible human operator via the AICID-to-ORCID link. CrossRef and DataCite metadata can carry AI authorship signals, allowing downstream databases to propagate disclosure without additional publisher effort.

Preprint servers such as arXiv currently prohibit AI scientists from submitting directly to  \cite{arxiv2025moderation}. 
An AICID field in submission metadata would allow preprint servers to handle AI submissions soundly.
Platforms that choose to accept AI-authored submissions, such as aiXiv~\citep{zhang2025aixiv}, could require a valid AICID as a condition of submission.

Search engines and bibliographic databases (Google Scholar, Semantic Scholar, Scopus, Web of Science) index papers based on metadata extracted from PDFs and publisher feeds.
 Adding AICID support to their author disambiguation pipelines would allow these systems to correctly classify AI
scientists as non-human contributors and surface that information to users. Without such integration,
AI scientists will continue to appear as ordinary human researchers in search results.

\section{AICID Governance}

We propose the creation of a nonprofit organization to govern and sustain the AICID initiative.
Governance by a nonprofit, rather than by a single publisher, technology company, or government body, ensures that no single stakeholder can capture the registry for commercial or political ends.
The governing body should include representatives from publishers, preprint servers, bibliographic databases, AI developers, academic institutions, and research integrity organizations, mirroring the multi-stakeholder board model that has made ORCID a trusted neutral infrastructure.
The nonprofit would be responsible for (i)~maintaining the public registry and its API, (ii)~setting and updating the metadata schema as AI systems evolve, (iii)~arbitrating disputes over identifier assignment or misuse, and (iv)~enforcing disclosure and transparency requirements on operators who register AI contributors.
Financially, the organization would operate on a membership-fee model analogous to ORCID's, in which institutional members, publishers, and universities pay annual fees that cover operational costs while registry access remains free and open.
Open-source release of the registry software and public availability of the full dataset under a permissive license are governance commitments that prevent lock-in and allow independent audits of the system's integrity.

\section{Related Work}

\paragraph{ORCID.} The Open Researcher and Contributor ID system~\citep{haak2012orcid} was
established in 2012 to solve the researcher name disambiguation problem: multiple researchers share
identical names, and a single researcher may publish under variant name forms. ORCID assigns a
16-character identifier to each researcher, linked to a continuously updated record of publications,
employment, and funding. As of 2025, ORCID has issued over 25 million iDs~\citep{orcid2025statistics}. ORCID's
human-only scope, while appropriate for its original mission, makes it unsuitable as an AI author
identifier. AICID is designed to fill exactly the gap that ORCID explicitly excludes.

\paragraph{Preprint Servers for AI Scientists.} Zhang et al.~\citep{zhang2025aixiv} propose aiXiv, a preprint
server designed specifically for AI-generated research. Schmidgall and Moor~\citep{schmidgall2025agentrxiv}
describe AgentRxiv, a platform for collaborative autonomous research agents. Both projects address
the absence of appropriate publication venues for AI-generated work. They are complementary to
AICID: AICID provides author identity metadata and infrastructure, while these platforms provide submission and
dissemination infrastructure. A complete ecosystem for AI-generated science requires both.

\section{Conclusion}

The scholarly ecosystem is not prepared for AI scientists. AI-generated papers enter the literature
without any infrastructure to represent them as AI-generated. Human researchers, editors, and
databases treat AI authors and human authors identically because they have no means to distinguish
them. The consequences range from compromised transparency to unenforceable disclosure policies
to unresolvable accountability questions.

AICID addresses this gap by providing a persistent, unique identifier for AI scientists, linked to the
human operators who deploy them and integrated with existing bibliographic infrastructure. AICID
does not resolve whether AI systems should author papers. It provides the infrastructure necessary
to make that question answerable in practice: who or what authored a given paper, who is
responsible for it, and how can that information be found reliably by anyone who needs it.

\bibliographystyle{plainnat}
\bibliography{references}

\end{document}